\documentclass[twocolumn,superscriptaddress,showpacs,showkeys,preprintnumbers,amsmath,amssymb,aps]{revtex4-1}

\usepackage{graphicx} 
\usepackage{float}
\usepackage{units}
\usepackage{hyperref}
\usepackage{color}
\usepackage{booktabs}

\begin{document}

\preprint{Gasi \textit{et al.}, Fe$_{2}$YZ}

\title{New iron-based Heusler compounds Fe$_{2}$YZ: Comparison with theoretical predictions of
the crystal structure and magnetic properties}

\author{T.~Gasi}
\affiliation{Institut f{\"u}r Anorganische und Analytische Chemie,
             Johannes Gutenberg-Universit{\"a}t, 55128 Mainz, Germany}
\author{V.~Ksenofontov}
\affiliation{Institut f{\"u}r Anorganische und Analytische Chemie,
             Johannes Gutenberg-Universit{\"a}t, 55128 Mainz, Germany}
\author{J.~Kiss}
\affiliation{Max-Planck-Institut f\"ur Chemische Physik fester Stoffe, 01187  Dresden, Germany}
\author{S.~Chadov}
\affiliation{Max-Planck-Institut f\"ur Chemische Physik fester Stoffe,
  01187  Dresden, Germany}
\author{A.~K.~Nayak}
\affiliation{Max-Planck-Institut f\"ur Chemische Physik fester Stoffe,
  01187  Dresden, Germany}
  \author{M.~Nicklas}
\affiliation{Max-Planck-Institut f\"ur Chemische Physik fester Stoffe,
  01187  Dresden, Germany}
\author{J.~Winterlik}
\affiliation{Institut f{\"u}r Anorganische und Analytische Chemie,
             Johannes Gutenberg-Universit{\"a}t, 55128 Mainz, Germany}
\author{M.~Schwall}
\affiliation{Institut f{\"u}r Anorganische und Analytische Chemie,
             Johannes Gutenberg-Universit{\"a}t, 55128 Mainz, Germany}
\author{P.~Klaer}
\affiliation{Institut f{\"u}r Physik,
             Johannes Gutenberg-Universit{\"a}t, 55099 Mainz, Germany}
\author{P.~Adler}\email{adler@cpfs.mpg.de}
\affiliation{Max-Planck-Institut f\"ur Chemische Physik fester Stoffe,
  01187  Dresden, Germany}
\author{C.~Felser}\email{felser@cpfs.mpg.de}
\affiliation{Institut f{\"u}r Anorganische und Analytische Chemie,
             Johannes Gutenberg-Universit{\"a}t, 55128 Mainz, Germany}
\affiliation{Max-Planck-Institut f\"ur Chemische Physik fester Stoffe,
  01187  Dresden, Germany}

\date{\today}

\begin{abstract}

The present work reports on the new soft ferromagnetic Heusler phases Fe$_{2}$NiGe,
Fe$_{2}$CuGa, and Fe$_{2}$CuAl, which in previous theoretical studies have 
been predicted to exist in a tetragonal Heusler structure. 
Together with the known phases Fe$_{2}$CoGe and Fe$_{2}$NiGa these materials 
have been synthesized and characterized by powder XRD, $^{57}$Fe M{\"o}ssbauer 
spectroscopy, SQUID, and EDX measurements. In particular M{\"o}ssbauer spectroscopy
was used to monitor the degree of local atomic order/disorder and to estimate 
magnetic moments at the Fe sites from the hyperfine fields. 
It is shown that in contrast to the previous predictions all the materials 
except Fe$_{2}$NiGa basically adopt the inverse cubic Heusler (X-) structure 
with differing degrees of disorder. 
The experimental data are compared with results from {\it ab-inito} electronic 
structure calculations on LDA level incorporating the effects of atomic disorder
by using the coherent potential approximation (CPA). 
A good agreement between calculated and experimental magnetic moments is found 
for the cubic inverse Heusler phases. 
Model calculations on various atomic configurations demonstrate that antisite 
disorder tends to enhance the stability of the X-structure.

\end{abstract}



\maketitle
\section{Introduction}
Magnetic Heusler compounds with the general formula X$_{2}$YZ 
where X and Y are transition metal elements and Z is a main group element are
one of the most fascinating classes of modern magnetic materials, from both
a fundamental as well as an application-oriented point of view~\cite{GPF11,TRU10}. 
The large variety of physical properties which can be realized in Heusler-type 
materials is a consequence of their peculiar crystal structure and their great 
chemical flexibility. The basic cubic $L2_1$ crystal structure of X$_2$YZ phases
(space group $Fm\overline{3}m)$) is composed of four interpenetrating fcc sublattices
(see Fig.\ref{fig:Heusler_inv_Heusler}). 
It may be considered as a zincblende-like arrangement of Z and X atoms with the remaining 
X atoms in the tetrahedral and the Y atoms in the octahedral voids. 
This leads to a NaCl-type arrangement of the Y and Z atoms where each Y atom 
is surrounded by six Z atoms in the second coordination sphere, whereas the 
nearest neighbors of each Y atom are eight X atoms.
On the other hand, each X atom is surrounded by four Y and four Z atoms. 
An early contribution to the rationalization of the magnetic properties of Heusler 
compounds is the theoretical work by K\"ubler et al. on X$_2$MnZ-type 
compounds~\cite{KUB83}, which describes how large localized magnetic moments in 
the Mn sublattice are realized by an itinerant electronic system composed of 
hybridized Mn~d-X~d bands~\cite{KUB83}.
K{\"u}bler et al. emphasized the importance of indirect Mn-Z-Mn interactions 
for the magnetic properties. The various magneto-responsive effects of Heusler compounds such as 
magnetic-field-induced shape memory/strain effects, magnetoresistance, and 
magnetocaloric effects~\cite{EWF12,ADA06} promise a wide range of applications. 
For instance, ferromagnetic martensites are experimentally found in different 
half-Heusler alloys such as NiMnGa, NiFeGa, CoNiGa and related materials~\cite{YChY03}. 
All these effects occur in temperature ranges which are convenient for device operation.
The great current interest in Heusler compounds arises, however, at least partly from 
the observation of half-metallic ferromagnetism in several of these 
materials~\cite{FEL07,GPF11} as half-metallic ferromagnets have a great potential 
in the emerging field of spintronic applications~\cite{WFB11,MJG11}. 
Since Wurmehl {\it et~al.} have reported an exceptionally high magnetic moment
of 6~$\mu_{\rm B}$
and a Curie temperature $T_C$ of 1100~K for the half-metallic
Co$_{2}$FeSi~\cite{SGH06}, the interest for Heusler compounds has highly
increased, and the Co-based Heusler compounds have been studied
extensively~\cite{TRU10}. The combination of Co and Fe is particularly good for obtaining high $T_C$'s.\\
\indent
Not only the cubic but also tetragonally distorted Heusler compounds 
--like Mn$_3$Ga~\cite{WBF08}-- have found considerable attention as they are
attractive candidates for spin-transfer-torque applications~\cite{HZK12}. 
In this respect it is remarkable that in course of a recent theoretical 
study on potentially new Heusler compounds~\cite{GD08,GD09} several 
iron-based Fe$_2$YZ materials  with a regular tetragonal Heusler structure
have been predicted~\cite{GD09}. 
Actually it was found for these phases that the inverse cubic Heusler structure
(X-structure, space group $F\overline{4}3m)$), should be more stable than the 
regular one (see Fig.\ref{fig:Heusler_inv_Heusler}). 
The X-structure may be formally written as (FeY)FeZ where the Fe atoms occupy 
the octahedral and tetrahedral voids of the zincblende-type lattice to equal parts.
The inverse Fe$_2$YZ structures in some of the materials are, however, predicted
to be unstable towards a tetragonal distortion which tends to maximize the 
bonding interactions~\cite{GD09}. Motivated by the potential importance of tetragonal iron-based Heusler compounds
for applications, especially as potential rare-earth free hard magnets, and by the fundamental question of electronic instabilities in 
Heusler compounds this work reports the synthesis and characterization of the 
new compounds Fe$_2$NiGe, Fe$_2$CuGa, and Fe$_2$CuAl, the existence of which 
has been predicted in Ref~\cite{GD09}.
In addition the properties of the known phases Fe$_2$CoGe~\cite{REN10} and 
Fe$_2$NiGa~\cite{JR00} have been investigated in more detail. 
Usually first-principle electronic structure calculations are based on a perfect crystal. 
However, Heusler compounds are ternary or even quaternary compounds which are widely
amenable to chemical disorder. 
This can be the reason for discrepancies between theoretical predictions and 
experimental results. 
This aspect turns out to be of crucial importance also for the present Fe$_2$YZ
materials and therefore will be discussed in detail. 
As X-ray diffraction (XRD) techniques are frequently insufficient for extracting
the detailed atomic order in Heusler materials additional experimental methods 
are required. 
Anomalous XRD and extended x-ray absorption fine structure (EXAFS) experiments
are versatile techniques for this purpose~\cite{OUA11}, while x-ray magnetic 
circular dichroism (XMCD) studies give access to site-selective magnetic moments~\cite{MEI11,KLA11}. 
In case of the present iron-based materials we have used
$^{57}$Fe-M{\"o}ssbauer spectroscopy as a simple laboratory-technique to 
characterize both the local atomic and magnetic order in our samples. 
The additional theoretical analysis based on first-principles electronic structure calculations treating  the
chemical disorder by means of the CPA method~\cite{Sov67,But85}(coherent potential
approximation), shows that the cubic structure of Fe$_2$YZ Heusler
materials becomes indeed  more stable when assuming certain degree of
anti-site disorder as suggested by the experiments. 


\section{Synthesis and characterization}
\subsection{Experimental details}
\label{sec:ed}
Fe$_{2}$YZ (Y=Co, Ni, Cu; Z=Al, Ga, Ge) compounds  were synthesized by repeated
arc-melting or ball milling of stoichiometric amounts of pure metals in an argon 
atmosphere under $10^{-4}$~mbar pressure. 
To ensure an oxygen-free atmosphere, titanium was used as getter material. 
The samples were  three times melted and turned over. 
The weight loss after the whole process was less than 0.5\%. 
Additionally, they were annealed in evacuated glass quartz tubes for two weeks 
at 673~K and then in order to test the influence of the annealing temperature
on the degree of disorder some of them were further annealed at 1173~K.
After annealing they were quenched in a mixture of ice and water. 
In addition it has been attempted to synthesize Fe$_2$ZnAl in a similar way, 
however, the Heusler phase could not be stabilized. 
The crystal structures of the products at room temperature were investigated by 
means of powder XRD using excitation by monochromatic 
Cu-K$_{\alpha1}$ ($\lambda_{\rm {Cu},K{\alpha1}}$=1.540598~{\AA}) or
Mo-K$_{\alpha 1,2}$ radiation ($\lambda_{\rm {Mo},K{\alpha1}}$=0.7093165~{\AA}) in 
$\theta$-$\theta$ scanning mode. 
Disc-shaped samples were used for Fe$_{2}$CoGe, Fe$_{2}$NiGa, Fe$_{2}$CuGa and Fe$_{2}$CuAl. 
Since Fe$_{2}$NiGe samples were better produced by ball-milling, they were measured as powders.
The XRD patterns were fitted by the FullProf software package~\cite{Coe07}, 
in case of Mo-K$_{\alpha 1,2}$ radiation the $\alpha_1$/$\alpha_2$ splitting of 
the radiation was taken into account.
The magnetization was measured by different SQUID magnetometers 
(MPMS-XL5 and MPMS SQUID VSM, Quantum Design) within the temperature range 1.8 to 950~K.
$^{57}$Fe-M{\"o}ssbauer measurements were performed in transmission, backscattering
and CEMS (conversion electron M{\"o}ssbauer spectroscopy) modes using a constant 
acceleration spectrometer with a $^{57}$Co~(Rh-matrix) source with  $h\nu=14.4$~keV. 
The backscattering spectra were recorded with the miniaturized M{\"o}ssbauer
spectrometer MIMOS~\cite{KLI11}.
Information about bulk properties was obtained from the transmission or 
backscattering spectra 
(information depth d~$\sim10^3$~{\AA}), while the CEM spectra
(d~$\sim10^2$~{\AA}) is more surface sensitive. 
To obtain the reliable distribution of hyperfine parameters the data were fitted 
using the Voigt-based fitting (VBF) model within the Recoil software
package~\cite{Coe08}.\\
\indent
The homogeneity and stoichiometry of the samples was controlled by a scanning
electron microscope (SEM, Jeol~JSM-6400) equipped with an
energy-dispersive X-ray (EDX) spectroscopy detection system (EUMEX~EDX). 
The measurements were carried out at $3\times10^{-6}$~mbar pressure. 
The acceleration voltage of 20~kV with inspection angle of 
35$^{\circ}$ was used. For the correction of the quantitative data the
so-called ZAF method was applied, which relies on atomic number (Z),
and on absorption (A) and fluorescence (F) effects. 
The images were acquired via the Digital Image Processing System (DIPS) 
and the quantitative chemical analysis was performed with the WINEDS~4.0 program.
According to the EDX analysis the Fe$_{2}$YZ alloys show a rather good 
homogeneity in their composition, except Fe$_{2}$NiGe (see Tab.~\ref{tab_EDX}).
\begin{table}
\setlength\belowcaptionskip{0.0cm}
\setlength\abovecaptionskip{0.1cm}
\begin{ruledtabular}
\begin{tabular}{c|ccc}
Compound          & $Fe$ & $\rm {Y}$ & $\rm {Z}$\\
\hline                            
Fe$_{2}$CoGe      &50.43   &25.36   &24.22 \\
Fe$_{2}$CuGa      &49.77   &25.94   &24.30 \\
Fe$_{2}$NiGa      & 50.78  &24.65   &24.57 \\
Fe$_{2}$NiGe      & 47.48  &27.55   &24.97 \\
Fe$_{2}$CuAl      & 48.32  &26.57   &25.11 \\
\end{tabular}
\end{ruledtabular}
 \caption{Relative concentration given as percentages of the samples 
 from the Fe$_{2}$YZ alloys according to an EDX analysis.
\label{tab_EDX}}
\end{table}
\subsection{XRD structural characterization}
\label{sec:struct}
\begin{figure}[t!]
   \centering
\setlength\belowcaptionskip{0.0cm}
\setlength\abovecaptionskip{0.1cm}
   \includegraphics[width=0.8\linewidth,clip]{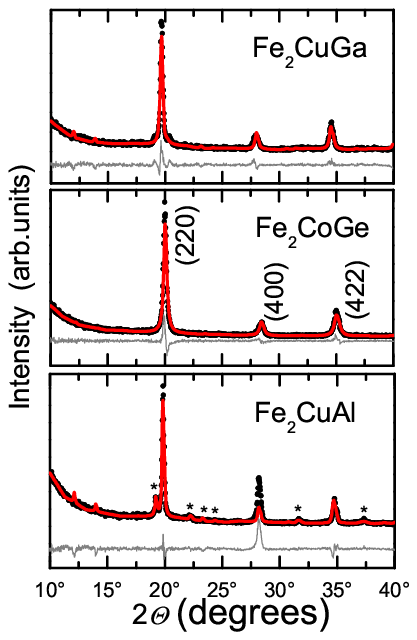}
   \caption{
(Color online)
XRD of Fe$_{2}$CuGa, Fe$_{2}$CoGe and Fe$_{2}$CuAl alloys annealed at 673~K. 
The diffraction patterns are compared to the results of Rietveld refinements.
Diffraction peaks marked with stars are assigned to an impurity phase Cu$_{x}$Al$_{y}$~\cite{MKB91}
(space group $Fm\overline{3}m$ with a = 3.63~{\AA}).
The measurements were carried out at RT using Mo-K$_{\alpha}$ radiation. 
\label{fig:Fe2CoGeundFe2CuAl_XRD}}
\end{figure}

\begin{figure}[t!]
   \centering
\setlength\belowcaptionskip{0.0cm}
\setlength\abovecaptionskip{0.1cm}
   \includegraphics[width=0.8\linewidth,clip]{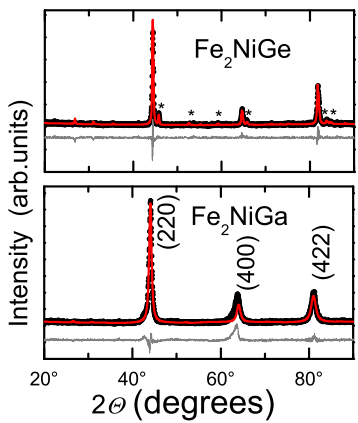}
   \caption{
(Color online)
XRD of Fe$_{2}$NiGe and Fe$_{2}$NiGa alloys annealed at 673~K.
The diffraction patterns are compared to the results of Rietveld refinements. 
Diffraction peaks marked with stars are assigned to an impurity phase Fe$_{x}$Ge~\cite{PNA78} 
(space group $P6_3/mmc$ with a = 3.97~{\AA} and c=5.05~{\AA}).
The measurements were carried out at RT using Cu-K$_{\alpha}$ radiation. 
\label{fig:XRD_400_NiGa_NiGe}}
\end{figure}

Typically well-ordered Heusler compounds crystallize in a cubic crystal structure,
either in the regular $L2_1$ (space group $Fm\overline{3}m)$) or in the inverse 
X-(space group $F\overline{4}3m)$) structure. 
For both structure types the XRD patterns exhibit additional (111) and (200) fcc 
superstructure reflections. 
In many cases the intensity ratio $I_{111}/I_{200}$ allows to distinguish them 
from each other. 
Based on plane-wave pseudopotential calculations, some of the present materials 
have been predicted to crystallize in a tetragonally distorted regular Heusler 
structure (space group $P4_2/ncm$)~\cite{GD09}).
The XRD patterns of the Fe$_2$YZ materials annealed at 673~K are shown in 
Fig.~\ref{fig:XRD_400_NiGa_NiGe} and in Fig.~\ref{fig:Fe2CoGeundFe2CuAl_XRD}. 
As is seen from the typical distances between the (220), (400), and (422) 
reflections (note that the Miller indices correspond to an fcc lattice) all 
materials crystallize in a cubic structure. 
Additional weak reflections in the patterns of Fe$_2$NiGe, Fe$_2$CuGa, and Fe$_2$CuAl 
point to the presence of a small fraction of impurity phases. 
However, none of the patterns indicates any features like peak splittings 
which could be attributed to a tetragonal distortion predicted for 
Fe$_2$NiGe, Fe$_2$CuGa, and Fe$_2$CuAl~\cite{GD09}. 
The lattice parameters obtained from Rietveld refinement of the data by assuming the inverse cubic Heusler structure 
(space group $F\overline{4}3m)$) are given in Table~\ref{exp_XRD}. 
\begin{figure}[t!]
   \centering
\includegraphics[width=0.45\linewidth,clip]{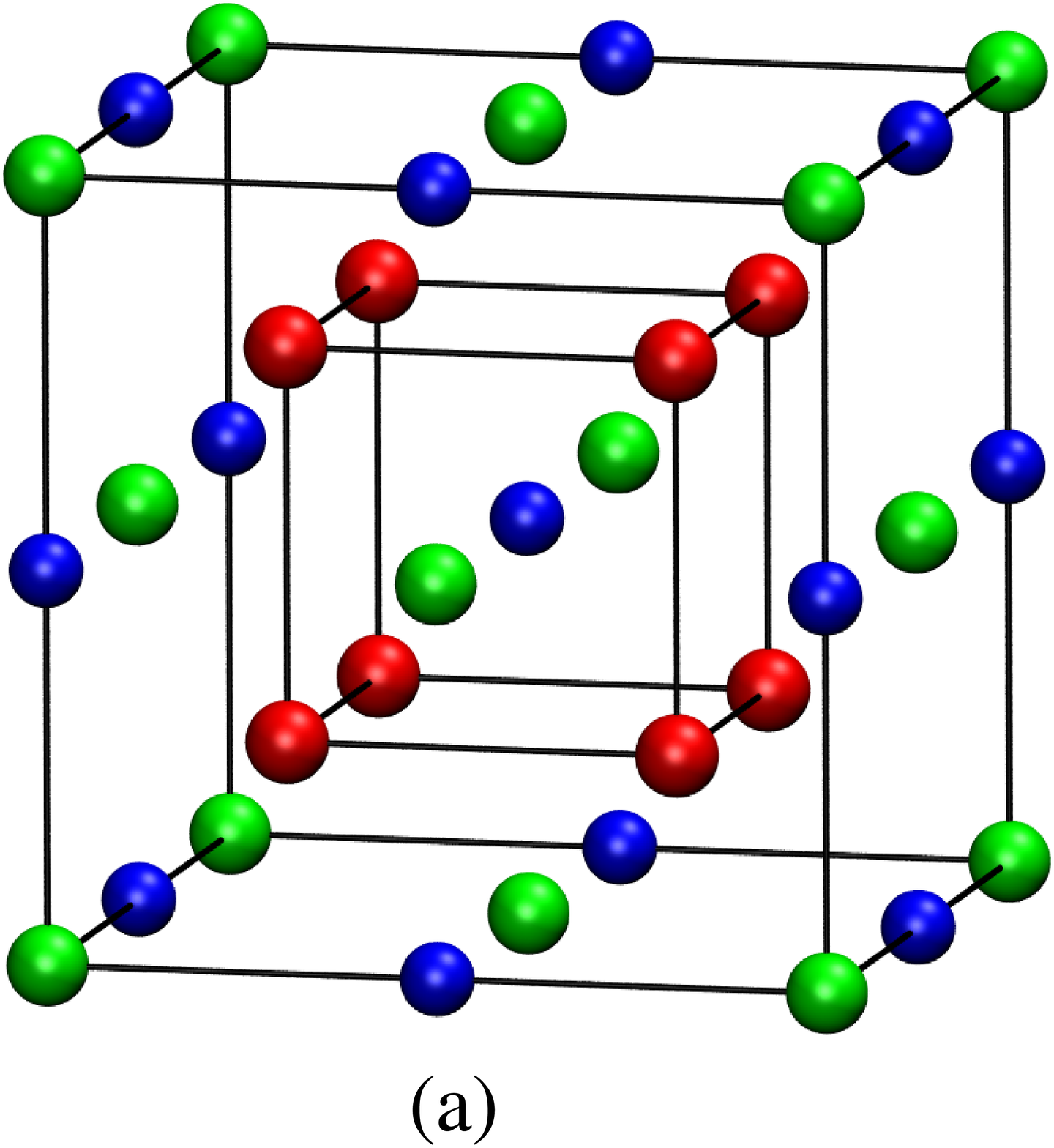}\hspace*{0.55cm}\includegraphics[width=0.45\linewidth,clip]{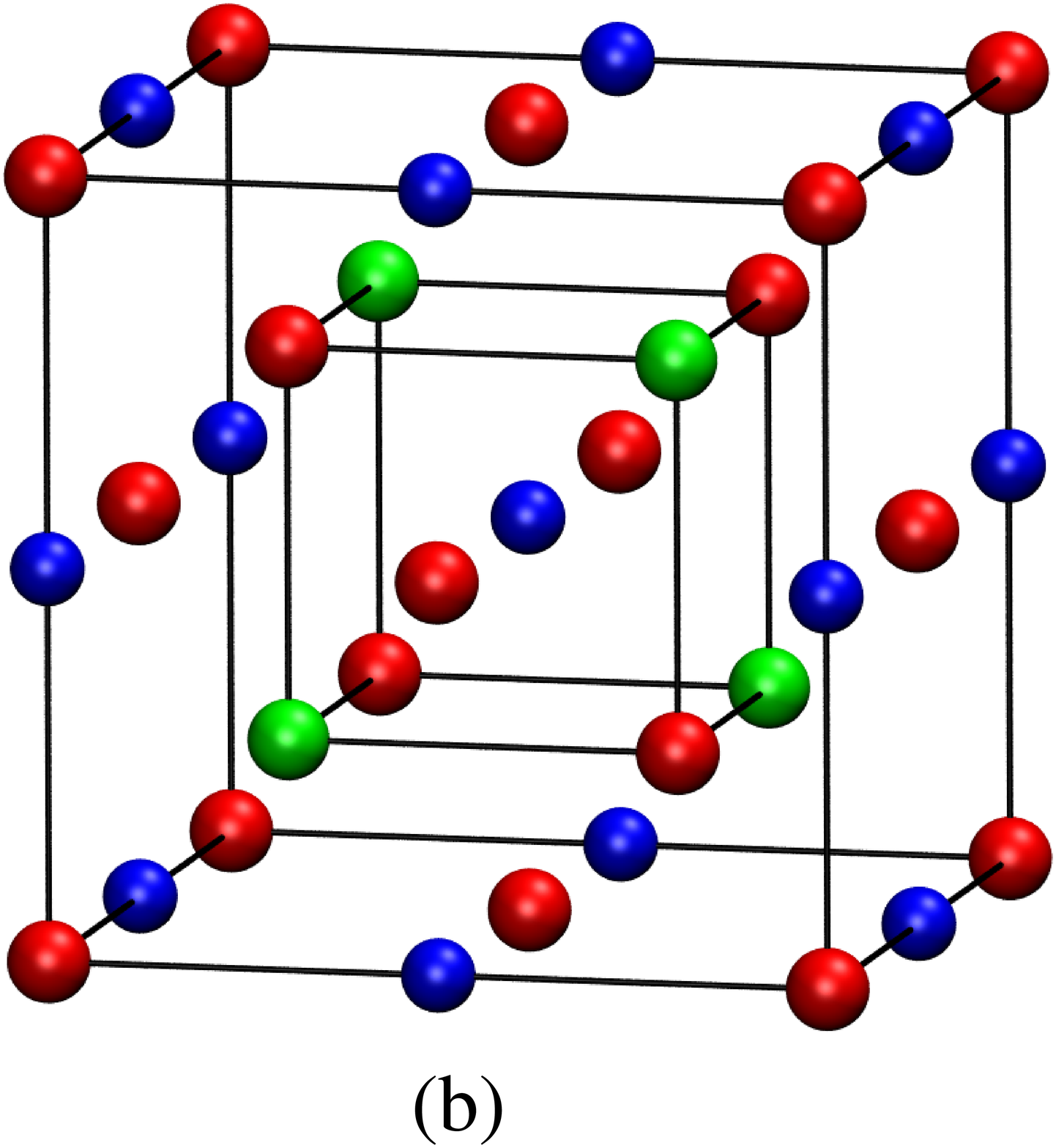}
   \caption{(color online) Conventional unit cell of the cubic Heusler (a) and inverse Heusler (b) X$_2$YZ compounds.
The X, Y, and Z atom types (where X=Fe for the materials studied in this paper) 
are shown with red, green, and blue spheres, respectively.
   \label{fig:Heusler_inv_Heusler}}
\end{figure}
\begin{figure}[t!]
   \centering
\setlength\belowcaptionskip{0.0cm}
\setlength\abovecaptionskip{0.1cm}
   \includegraphics[width=0.8\linewidth,clip]{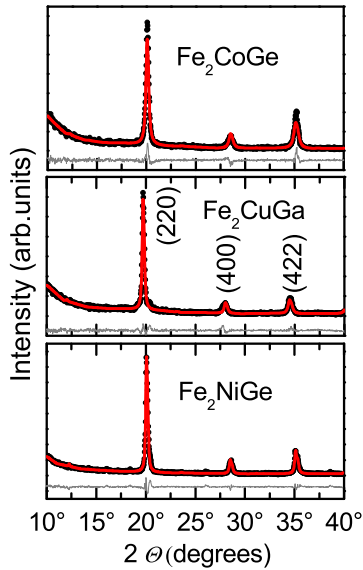}
   \caption{
(Color online)
XRD of Fe$_{2}$CoGe, Fe$_{2}$CuGa and Fe$_{2}$NiGe alloys annealed at 1173~K. 
The diffraction patterns are compared to the results of Rietveld refinements. 
\label{fig:XRD_900C}}
\end{figure}
\begin{table}
\setlength\belowcaptionskip{0.0cm}
\setlength\abovecaptionskip{0.1cm}
\begin{ruledtabular}
\begin{tabular}{c|cc}
Compound          & $a$ ({\AA})  & $a$ ({\AA})  \\
\hline                            
Fe$_{2}$CoGe      &5.78   &5.76    \\
Fe$_{2}$CuGa      &5.86   &5.86  \\
Fe$_{2}$NiGa      &5.81  &  -$^{\rm{a)}}$ \\
Fe$_{2}$NiGe      &5.76  &5.76   \\
Fe$_{2}$CuAl      &5.83  &   -$^{\rm{a)}}$ \\
\end{tabular}
\end{ruledtabular}
\caption{
The lattice parameters of Fe$_2$YZ were determined at room temperature and are given for an fcc lattice. 
Left are shown the lattice parameters of Fe$_2$YZ annealed at 673~K and 
right at 1173~K \label{exp_XRD}, and $^{a)}$ marks those cases which have involved decomposition of
the sample.}
\end{table}
The fcc superstructure reflections are not discernible in any of the XRD patterns.
Two reasons have to be considered for that.
First, the samples may be disordered according to the $A2$ type 
(with all lattice sites randomly occupied by the constituting elements). 
This results in a bcc-type ($Pm\overline{3}m$) diffraction pattern, where no 
superstructure reflections appear. 
Secondly, in all compounds except Fe$_2$CuAl the constituting elements are
entirely from the $4^{th}$ row of the periodic table, which results in similar
scattering factors. 
This leads to a virtual extinction of the (111) and (200) reflections, or at
least, to very low intensities, if the standard laboratory sources with 
Cu-K$_\alpha$ or Mo-K$_\alpha$ radiation are used. 
In fact for a well-ordered sample of Fe$_2$CoGe very weak superstructure reflections
have been observed~\cite{REN10}. 
Their detection requires, however, an improved signal-to-noise ratio of the XRD
patterns and thus considerably longer measurement times. 
For studying the influence of the annealing temperature on the sample quality,
in addition to the samples annealed at 673~K, the XRD patterns of samples 
annealed at 1173~K have been also recorded. 
In case of Fe$_2$NiGa and Fe$_2$CuAl sample decomposition was observed 
(XRD patterns not shown), the XRD patterns of the other samples (Fig.~\ref{fig:XRD_900C}) revealed only 
minor changes. 
The phase purity of Fe$_2$NiGe appears to be slightly improved by annealing at higher 
temperature, but the M{\"o}ssbauer spectra still revealed an impurity signal. 
Therefore, if not explicitly stated otherwise, we will focus in the
rest of the paper on the samples annealed at 673~K.\\
\indent 
In summary, the XRD patterns verify that the synthesized Fe$_2$YZ materials all 
crystallize in cubic Heusler-type structures. 
It is, however, impossible to derive the exact atomic order from the XRD data. 
More information about the atomic and magnetic order is obtained by 
$^{57}$Fe-M{\"o}ssbauer spectroscopy, which is a local probe technique.
The results obtained with this technique will be presented below.
\subsection{Characterization of magnetic properties}
\label{sec:magn}
\begin{figure}[t!]
\centering
\setlength\belowcaptionskip{0.0cm}
\setlength\abovecaptionskip{0.1cm}
   \includegraphics[width=0.8\linewidth,clip]{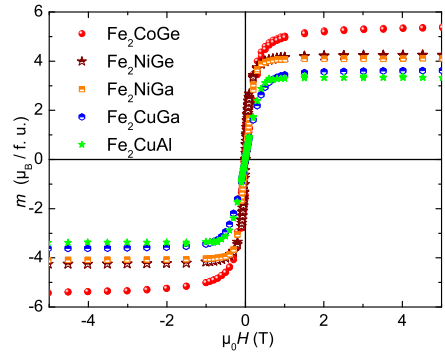}
   \caption{
(Color online)
Field-dependent magnetization $m$ per formula unit of Fe$_{2}$YZ compounds annealed at 673 K and
     measured at 5~K.\label{fig:Hysteresenbei5K_P}} 
\end{figure}

\begin{figure}[t!]
\centering
\setlength\belowcaptionskip{0.0cm}
\setlength\abovecaptionskip{0.1cm}
   \includegraphics[width=0.8\linewidth,clip]{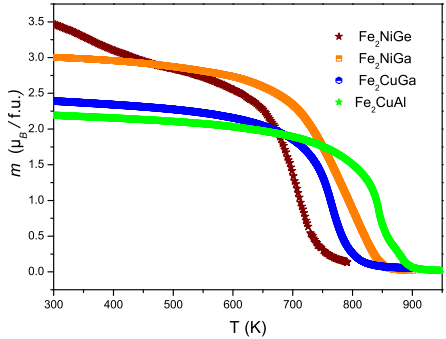}
   \caption{
(Color online)
Temperature-dependent magnetization $m$ per formula unit of Fe$_{2}$YZ alloys annealed at 673~K.
            The measurements were performed in the range of 300--900~K
             at 0.1~T (for Fe$_{2}$CuAl, Fe$_{2}$CuGa, Fe$_{2}$NiGa, Fe$_{2}$CoGe) and 
             1~T (for Fe$_{2}$NiGe) induction fields.
\label{fig:HT_Fe2YZ} }
\end{figure}

As displayed in Fig.~\ref{fig:Hysteresenbei5K_P}, all compounds exhibit soft ferromagnetism. 
It is noted, that the magnetic behavior slightly depends on the annealing temperature. 
The isothermic magnetization curves at 5~K are essentially saturated in an 
induction field of 5~T. 
The magnetic moments at 5~K and 5~T together with the magnetic ordering temperatures 
$T_C$ are summarized in Table~\ref{tab:magnet}. 
The latter have been derived from the drop in the temperature dependent 
magnetization curves measured at 0.1~T or 1~T (~\ref{fig:HT_Fe2YZ}). 
Additional anomalies in the T-dependence of Fe$_2$NiGe and Fe$_2$CuAl point to
the ordering of a minority phase in these samples. 
The valence electron concentration (VEC) per formula unit and magnetic 
moments obtained from electronic structure calculations are included in 
Table~\ref{tab:magnet}, too. 
All materials considered in this paper contain 29 or 30 valence electrons per formula unit. 
The magnetic moment of the 673~K-annealed sample of Fe$_2$CoGe is somewhat larger
than the value of 5.1~$\mu_{\rm B}$ reported earlier~\cite{REN10}, while that of the 
1173~K-sample is in better agreement with the published data. 
The experimental magnetic moments of Fe$_2$CoGe are quite close to the value of 
5~$\mu_{B}$ expected from the generalized Slater-Pauling (SP) rule. 
This suggests that Fe$_2$CoGe could be a half-metallic ferromagnet,
however, the spin polarization appears to be not complete~\cite{REN10}(see also Section III). 
The magnetic moments of all the other materials do not follow the generalized SP rule. 
The magnetic ordering temperatures are quite high \mbox{($T_C$ $>$ 700~K)} which makes
the materials suitable for potential applications. A detailed comparison of experimental magnetic
moments from both bulk magnetization and M{\"o}ssbauer measurements with theoretical moments will be
compiled in Section~III.

\begin{table*}
\caption{Summary of the magnetic properties (magnetic moments $m$, Curie temperatures T$_{c}$)
of Fe$_{2}$YZ alloys obtained from experiment (SQUID and M\"ossbauer measurements, 
left part of the Table) and from {\it ab-initio} electronic structure calculations 
(right part of the Table, indicated with $m_{\rm{th}}$). 
The total magnetic moments $m$ per formula unit are obtained from magnetization measurements at
5~K, whereas the magnetic moments at the Fe sites are derived from room 
temperature M{\"o}ssbauer spectra.
\label{tab:magnet}}
\smallskip
\centering
\begin{ruledtabular}
 \begin{tabular}{ c| c c c c c c|| c c c}
 compound & VEC\hspace{1pt}$^{\rm{a)}}$	&  $m^{b)}$   &  T$_{c}^{\rm{b)}}$
& $m^{\rm{c)}}$    & $m$~[FeA]$^{\rm{d)}}$ & $m$[FeB]$^{\rm{d)}}$ &
$m_{\rm{th}}^{\rm{e)}}$ &  $m_{\rm{th}}^{\rm{e)}}$[FeA] & $m_{\rm{th}}^{\rm{e)}}$[FeB] \\
  &           & ($\mu_{B}$) &(K) & ($\mu_{B}$) & ($\mu_{B}$)       & ($\mu_{B}$)  	& ($\mu_{B}$)     & ($\mu_{B}$)      & ($\mu_{B}$)	\\
\hline

Fe$_{2}$CoGe	&29& 5.40&925  & 5.0   &2.6     & 1.6         &5.15       & 2.68   & 1.51      \\ 
Fe$_{2}$NiGe    &30& 4.29&750  &4.46   &2.5     & 1.7         &4.47      & 2.53   &1.67        \\ 
Fe$_{2}$NiGa 	&29& 4.20& 845 & -     & -      & 2.05$^{\rm{f)}}$        &4.81       &2.64    &1.82      \\ 
Fe$_{2}$CuGa	&30& 3.60&798  & 3.40  & 2.5    & 1.7         & 4.04      & 2.34   & 1.78      \\ 
            	&  &     &     &       &        & 2.1$^{\rm{g)}}$    &           &        &    \\
Fe$_{2}$CuAl  &30& 3.30& 875 & -     & 2.2      & 1.7         & 3.80      & 2.23    &1.67        \\
\end{tabular}
\end{ruledtabular}
\begin{flushleft}
a) VEC is the valence electron concentration\\
b) measurement for the 673~K-annealed sample\\
c) measured for the sample annealed at 1173~K\\
d) estimated from the M{\"o}ssbauer hyperfine fields of the 673~K-annealed sample at RT\\
e) theoretical moments for the fully ordered inverse cubic Heusler phase
calculated for the experimental lattice parameters\\
f) derived from average H$_{\rm{hf}}$ value of the M{\"o}ssbauer hyperfine pattern\\ 
g) derived from the average H$_{\rm{hf}}$ value of the broad component in the
M{\"o}ssbauer hyperfine pattern
\end{flushleft}
\end{table*}

\subsection{M{\"o}ssbauer measurements}
\label{sec:moes}
%
\begin{figure}[t!]
\centering
\setlength\belowcaptionskip{0.0cm}
\setlength\abovecaptionskip{0.1cm}
\includegraphics[width=1.0\linewidth,clip]{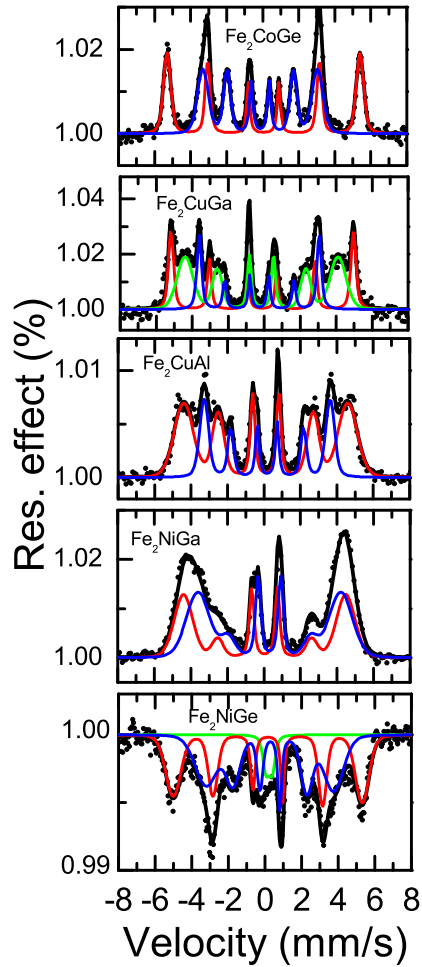}
\caption{
(color online)
M{\"o}ssbauer spectra of the Fe$_{2}$YZ samples annealed at 673~K. The spectrum of Fe$_{2}$NiGe 
is a transmission spectrum, the others are backscattering spectra.
The experimental points are shown as black dots, fitted by the black solid line 
(total signal). The total signal is decomposed into two (or three) distinct 
contributions (red, blue and green, respectively), each corresponding to a 
certain environment of the Fe atoms. 
\label{fig:Moss_400}}
\end{figure}
%
\begin{figure}[t!]
\centering
\setlength\belowcaptionskip{0.0cm}
\setlength\abovecaptionskip{0.1cm}
\includegraphics[width=0.8\linewidth,clip]{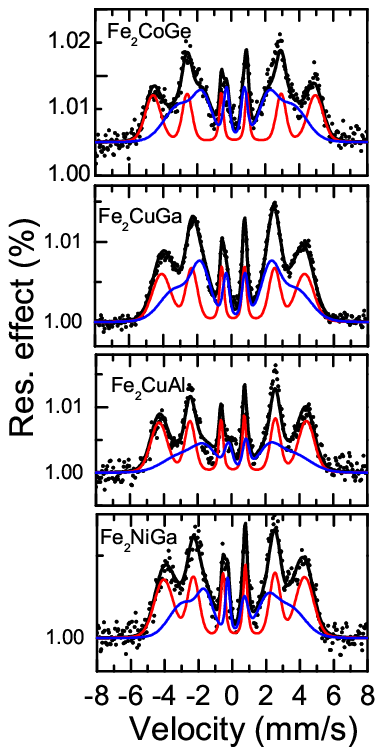}
\caption{
(color online)
Conversion Electron M{\"o}ssbauer spectra of Fe$_{2}$YZ samples annealed
at 673~K and measured at RT. The experimental points are shown as black dots, 
fitted by the black solid line (total signal). The total signal is decomposed into
two distinct contributions (red and blue, respectively), each corresponding to a certain
environment of the Fe atoms. 
\label{fig:Fe2YZ_CEMS}}
\end{figure}
The potential of M{\"o}ssbauer spectroscopy for studying the degree of atomic order
in iron-containing Heusler compounds has been demonstrated previously for the series
Co$_{2-x}$Fe$_{1-x}$Si~\cite{VBC08} and Co$_{2}$Mn$_{1-x}$Fe$_{x}$Al~\cite{VBC09}.
The former compounds crystallize in well-ordered cubic Heusler structures which 
evolve from the regular $L2_{1}$ structure for Co$_{2}$FeSi to the inverse X 
structure for Fe$_{2}$CoSi. 
The M{\"o}ssbauer spectra of Co$_{2-x}$Fe$_{1-x}$Si $(0.1 \leq x \leq 0.9)$ are 
composed of two sharp six-line patterns with magnetic hyperfine fields $H_{\rm hf}$
in the range 320--340~kOe (sextet A) and 185--195~kOe (sextet B), respectively.
The successive replacement of Co by Fe atoms in Co$_{2-x}$Fe$_x$Si leads to a 
corresponding increase in the area fraction of sextet B. 
From the above mentioned series it can be concluded unambiguously that the A sextet corresponds 
to Fe sites with (8-4x) Co and 4x Fe atoms in the first coordination sphere, 
the B sextet corresponds to Fe sites with four Fe(A) and four non-magnetic Si 
neighbors, independent of x. 
The variation in the $H_{\rm hf}$ values of the two sites due to the substitution
is quite small. 
In the $L2_1$ structure of Co$_2$FeSi the Fe(A) atoms have $O_h$ symmetry with
eight Co nearest neighbors and six Si atoms in the second coordination shell.
In the final member compound Fe$_2$CoSi all atoms have $T_d$ symmetry and the Fe(A) sites
are surrounded by four Fe(B) and four Co atoms. 
On the other hand a B2-type disorder in the series Co$_2$Mn$_{1-x}$Fe$_x$Al 
gives rise to a strongly broadened hyperfine pattern which reflects the 
superposition of Fe sites in different coordination environments. 
A detailed analysis of these spectra has been given in Refs.~\onlinecite{VBC08,KWW10}.
The M{\"o}ssbauer spectra demonstrating the hyperfine field distribution at RT 
for the present Fe$_2$YZ materials annealed at 673~K are shown in
Fig.~\ref{fig:Moss_400} and~Fig.~\ref{fig:Fe2YZ_CEMS}. 
The transmission and backscattering spectra in Fig.~\ref{fig:Moss_400}  are 
representative for the bulk materials and will be discussed in detail below. 
The more surface sensitive conversion electron M{\"o}ssbauer (CEM) spectra (Fig.~\ref{fig:Fe2YZ_CEMS}) reveal additional
line broadening and the details seen in the backscattering or transmission spectra are lost. This points to enhanced disorder in the surface layers. 
Also M{\"o}ssbauer spectra of the samples annealed at 1173~K were recorded. 
As there was no improvement of atomic order in any of the samples and according
to XRD even partial decomposition occurred for Fe$_2$NiGa and Fe$_2$CuAl, we 
restrict the discussion to the 673~K-annealed samples. 
All data were fitted with a Voigt profile model which yields a distribution of
hyperfine fields. 
Actually this approximates the superposition of local coordination environments
as has been exemplified in more detail in Ref~\onlinecite{VBC09}.
\begin{table}
\setlength\belowcaptionskip{0.0cm}
\setlength\abovecaptionskip{0.1cm}
\caption{M{\"o}ssbauer data of the annealed samples at 673~K of the Fe$_{2}$YZ
compounds. $H_{\rm hf}$ denotes the hyperfine field, IS the isomer shift, and A the relative area fraction. The maximum errors in $H_{\rm hf}$,A, and IS were 
$\pm$ 4 kOe, $\pm$ 3{\%}, and $\pm$ 0.02 mm/s, respectively.}
\smallskip
\centering
\begin{ruledtabular}
\begin{tabular}{c|ccc} 
& \multicolumn{3}{c}{$T_{\rm anneal}=673$~K}  \\
\hline
\scriptsize sample & \scriptsize $H_{\rm hf}$~(kOe) & \scriptsize
$A$~(\%) & \scriptsize $IS$~(mm/s) \\
\hline
Fe$_{2}$CoGe            & 332                &48        &0.05               \\
                        & 197                &52        &-0.14               \\ 
Fe$_{2}$CuGa            & 314                &28        &-0.11             \\
                        & 264                &49        &-0.15              \\
                        & 207                &22        &-0.27            \\
Fe$_{2}$NiGa            & 275                &44        &0.04                \\
                        & 242                &56        &0.30              \\
Fe$_{2}$NiGe            & 320                &44        &0.15             \\
                        & 216                &51        &0.27              \\
Fe$_{2}$CuAl            & 279                &67        &0.10              \\
                        & 214                &33        &0.18              \\ 
\end{tabular}
\end{ruledtabular}
\label{tab_moessbauer}
\end{table}

The values of the M{\"o}ssbauer parameters of Fe$_{2}$YZ obtained from the data
evaluation for bulk samples are summarized in Table~\ref{tab_moessbauer}.
All spectra can be described by two or three hyperfine sextets which show
that the corresponding materials are magnetically ordered at room temperature.
The detailed spectral shapes differ, however, which suggests a varying degree 
of disorder within the series of compounds.  
The M{\"o}ssbauer spectra and  $H_{\rm hf}$ values of Fe$_2$CoGe are close to
those of Co$_{2-x}$Fe$_{1+x}$Si~\cite{VBC08} with two well-defined sextets. 
Thus, the sextet with higher $H_{\rm hf}$ is assigned to Fe($A$) sites and the
second sextet with lower $H_{\rm hf}$ value to the Fe($B$) sites. 
The intensity ratio of the two sextets is approximately 1:1. 
These data evidence that Fe$_2$CoGe crystallizes in the inverse
Heusler (X) structure like Fe$_2$CoSi. 
M{\"o}ssbauer spectra not only give information about the atomic distribution 
over the crystallographic sites but also allow an estimation of the ordered 
magnetic moments for Fe atoms at the A and B sites, respectively.
In cubic or nearly cubic compounds the internal magnetic hyperfine field is 
determined by the Fermi contact term, which arises from  the polarization of the s-electron
density at the nucleus. 
It has been shown that $H_{\rm hf}$ scales linearly with the magnetic moments 
determined from neutron or magnetization studies~\cite{MAS79}. 
Taking a slope of 125~kOe/$\mu_{\rm B}$ for Fe germanides from Ref.~\onlinecite{MAS79} one 
roughly estimates the magnetic moments as $\sim 2.6~\mu_{\rm B}$ and $\sim 1.6~\mu_{\rm B}$
for Fe(A) and Fe(B) atoms in Fe$_2$CoGe, respectively. 
Considering the total magnetic moment of 5.4~$\mu_{\rm B}$ from the magnetization 
studies (at 5~K) the remaining Co moment must be about 1.1~$\mu_{\rm B}$ in case of 
ferromagnetic ordering. 
These estimates are in fair agreement with the calculated moments of
2.7~$\mu_{\rm B}$ and 1.4~$\mu_{\rm B}$ for the Fe sites and 0.9~$\mu_{\rm B}$ for the Co sites reported
in Ref.~\onlinecite{REN10}. As for all the present materials the Curie temperatures are above 700 K the estimates of the moments from the room temperature M{\"o}ssbauer
spectra can be used for comparison with theoretical results. These data are included in Table III. 
Quite similar spectra are observed for Fe$_2$NiGe. 
Accordingly, also Fe$_2$NiGe essentially crystallizes in the inverse Heusler 
structure but line broadening for both subspectra and the small deviation from
the ideal intensity ratio of 1:1 indicate some disorder. 
In case of Fe$_2$NiGe, from the $H_{\rm hf}$ values (see Table~\ref{tab_moessbauer}) 
one estimates the magnetic moments as 2.5~$\mu_{\rm B}$ and 1.7~$\mu_{\rm B}$ for the Fe(A) 
and Fe(B) atoms, respectively. 
The sum of the moments nearly corresponds to the total magnetic moment of 
4.3~$\mu_{\rm B}$ for Fe$_2$NiGe obtained from magnetometry, which suggests that
the Ni sites do not contribute much to the magnetic ordering. 
It is noted, however, that the present material is not completely single-phase. 
The M{\"o}ssbauer spectrum reveals a 6\% contribution of a paramagnetic quadrupole doublet ($IS$ = 0.23 mm/s, $\Delta E_{\rm Q}$ = 0.37 mm/s).
In agreement with the clues from the XRD patterns this signal is assigned to an
Fe$_x$Ge$_y$ impurity as the magnetic behavior of this system is known to depend
strongly on the Fe-Ge ratio\cite{MAS79}.\\
\indent
In contrast to Fe$_2$CoGe and Fe$_2$NiGe the M{\"o}ssbauer spectrum of 
Fe$_2$NiGa reveals enhanced disorder. 
The two subspectra are no longer resolved and the spectra were modeled by a 
superposition of two broad sextets the average $H_{\rm hf}$ values of which 
are only slightly different. 
Hence, the distinction between well defined A and B sites is not meaningful. 
The disorder in Fe$_2$NiGa leads to a pronounced distribution of hyperfine 
fields and of the corresponding magnetic moments over various sites with 
differing local coordination environment. 
The nearly structureless spectra may point towards an A2-type disorder 
where alloying occurs between all sites. 
The average $H_{\rm hf}$ of 255~kOe roughly corresponds to an ordered moment of 
2~$\mu_{\rm B}$ per Fe or 4~$\mu_{\rm B}$ per formula unit which is close to the total 
moment of about 4.2~$\mu_{\rm B}$. 
The considerable disorder in Fe$_2$NiGa is also suggested by the XRD pattern 
which reveals a pronounced broadening of the (400) and (422) reflections.
Additional information about the orientation of the magnetic moments can be 
derived from the intensities of the Fe$_2$NiGa spectrum. 
As it is well known~\cite{ESS80} for $^{57}$Fe, in case of no or small quadrupole
interaction only 6 lines of the 8 possible transitions in a magnetic field are 
observed with the intensity ratio ${3:Z:1:1:Z:3}$. 
The $Z$ value is characteristic for the relative intensity of the 
{$\pm$$\nicefrac{1}{2}$~$\longrightarrow$~$\pm\nicefrac{1}{2}$} transitions and 
depends on the angle $\theta$ between the propagation vector of the 
$\gamma$-radiation and the direction of $H_{\rm hf}$: $0\leq Z\leq4$. 
For a random orientation (powder) and thin absorbers  $Z$ = 2, whereas Z = 4 for 
$\theta$ = 90$^\circ$ and Z = 0 for $\theta$ = 0$^\circ$, respectively. 
Preferred orientation of crystallites accordingly leads to deviation from the
ideal powder value $Z$ = 2. The backscattering spectrum of Fe$_2$NiGa 
(Fig.~\ref{fig:Moss_400}) reveals a strongly diminished relative intensity of 
the second and fifth lines which suggests that the magnetization occurs 
essentially along the direction of the $\gamma$-beam, i.e. perpendicular to the
sample surface.\\
\indent
Well ordered and disordered regions appear to coexist in the sample of Fe$_2$CuGa.
The spectra of Fe$_2$CuGa are best described by two sharp and an additional 
broad sextet. 
The $H_{\rm hf}$ values and thus also the magnetic moments of the sharp features
are very similar to those of Fe$_2$NiGe. 
These signals are attributed to Fe(A) and Fe(B) sites in well ordered domains of 
the material with X structure. 
The broad feature is assigned to domains with strong disorder. 
The average $H_{\rm hf}$ of ~265~kOe corresponds just to the average of the 
$H_{\rm hf}$ values of the A and B sites. 
Accordingly, although the distribution of the moments is changed in course of
the disorder the average magnetic moment (4.2~$\mu_{\rm B}$ per formula unit)
nearly remains the same. 
This is in agreement with the observation that the total magnetic moments in 
Heusler compounds frequently are not changed much by disorder. 
Similar spectral shapes are also obtained for Fe$_2$CuAl. 
The spectra have been approximated by two hyperfine sextets. 
The sharp sextet with  $H_{\rm hf}$=214~kOe is assigned to Fe(B) sites with 
regular coordination geometry (4~Fe and 4~Al neighbors). 
Its intensity contribution (33\%) is, however, smaller than for an ideal inverse 
Heusler phase. 
The broad second sextet thus accounts for a superposition of Fe sites with 
differing environment due to disorder. 
As in other Al containing Heusler compounds a pronounced B2-type exchange between Fe and Al 
sites is expected to contribute to the disorder~\cite{VBC09}. Antisite disorder between the Fe(B) and Al sites is possibly the origin for an enhanced intensity contribution of Fe(A)-like signals as Fe atoms at Al sites are in a similar environment as Fe(A) atoms. There is no clear evidence of an Fe-based impurity in the spectrum of Fe$_2$CuAl
although an impurity phase was detected in the XRD pattern. 
This is in accord with the clue from the XRD data that the impurity corresponds
to a Cu$_{x}$Al$_{y}$ phase.\\ 
\indent
Having established the typical features of the M{\"o}ssbauer spectra for the
present materials we now discuss the results with respect to the theoretical 
predictions of Ref.~\onlinecite{GD09}. 
For Fe$_2$CoGe the inverse cubic Heusler structure was obtained as the most stable one. 
This is in agreement with the M{\"o}ssbauer spectra of Fe$_2$CoGe which show two 
well-defined Fe sites. 
For all the other materials, however, the structure predictions are contradicted
by the M{\"o}ssbauer spectra. 
In Ref.~\onlinecite{GD09} an inverse Heusler structure was suggested for Fe$_2$NiGa. 
In contrast, the M{\"o}ssbauer spectrum of Fe$_2$NiGa does not reveal the typical 
pattern of the inverse Heusler structure. 
The broad features rather point towards a strongly disordered arrangement of the atoms. 
In case of Fe$_2$NiGe, Fe$_2$CuGa, and Fe$_2$CuAl the inverse Heusler structure 
was shown to be more stable than the regular one, but it was suggested that there
is an inherent tendency for a structural distortion. 
Hence, it was predicted that the regular tetragonal Heusler structure becomes 
more stable for these materials. 
As it has been shown already in Section~\ref{sec:struct}, there is no sign of a 
tetragonal distortion in any of the present XRD patterns. 
The M{\"o}ssbauer spectrum suggests that it is essentially the inverse Heusler 
structure which is adopted by Fe$_2$NiGe. 
Also the spectra of Fe$_2$CuGa and Fe$_2$CuAl show signatures of the inverse 
Heusler structure, however, pronounced disorder is apparent from the 
spectra too. Furthermore, there is no indication of quadrupole interactions in the M{\"o}ssbauer
spectra which would be expected for tetragonal Heusler phases.\\
\indent
Finally, we mention that we have tried to synthesize Fe$_2$ZnAl, which has also 
been predicted as a regular tetragonal Heusler phase~\cite{GD09} with a magnetic
moment of $\sim $4~$\mu_{\rm B}$. 
However, analysis of the M{\"o}ssbauer spectra and XRD data of a sample with nominal composition Fe$_2$ZnAl
revealed the presence of a non-magnetic component which was identified as FeAl. 
An additional magnetic component, the fraction of which increases from 10\% in 
the 673~K annealed sample to 55\% in the 1173~K sample presumably corresponds to Fe$_3$Al. 
Thus, the stabilization of a Heusler phase was not successful in this case.

\section{Results of the electronic structure calculations}

In order to gain more insight into the atomic and electronic structure and
magnetism of the Fe$_2$YZ compounds, we have performed {\em ab-initio}
electronic structure calculations based on density functional theory (DFT).
All calculations were carried out by using the fully-relativistic
Korringa-Kohn-Rostoker (KKR) Green's function method as implemented in the 
SPR-KKR package~\cite{EKM11}.
The exchange and correlation was treated by the Vosko-Wilk-Nusair form of the local density approximation (LDA)~\cite{VWN80}.\\
The lattice parameters were taken from the experimental data (see Table~\ref{exp_XRD}).
The primitive unit cell contains four atoms in the lattice with the Wyckoff positions 
\mbox{A (0, 0, 0)}, 
\mbox{B ($\nicefrac{1}{4}$, $\nicefrac{1}{4}$, $\nicefrac{1}{4}$)},
\mbox{C ($\nicefrac{1}{2}$, $\nicefrac{1}{2}$, $\nicefrac{1}{2}$)},
\mbox{D ($\nicefrac{3}{4}$, $\nicefrac{3}{4}$, $\nicefrac{3}{4}$)}.
In the regular Heusler structure the B and D positions are equivalent
due to inversion symmetry, whereas in case of the inverse Heusler structure these sites
become inequivalent (see Fig.~\ref{fig:Heusler_inv_Heusler}). 
In case of the fully ordered Fe$_2$YZ inverse Heusler compounds
sites A and B are occupied by Fe, which we will refer as Fe(A) and Fe(B) site,
respectively. Since the M{\"o}ssbauer measurements have clearly shown that the Heusler 
compounds in the Fe$_2$YZ series show certain degree of chemical disorder, 
we have considered  the anti-site disorder by using the so-called coherent potential approximation
(CPA)~\cite{Sov67,But85}.
In the following we will consider the cubic Heusler phases including the effects
of disorder and non-stoichiometry.

\subsection{Fe$_2$CoGe}
To check whether the methodology presented above is suitable for the
description of structural and magnetic properties of the Fe$_2$YZ materials,
first we have performed a set of calculations for Fe$_2$CoGe 
and compared the results with the data already available from the 
literature~\cite{REN10}. We have considered four different model systems 
in the cubic structure, where the site occupations were selected with the following configurations:
(a)~fully ordered Heusler structure;
(b)~chemically disordered structure (50~\% of Fe is mixed with Co);
(c)~chemically disordered inverse Heusler phase (50~\% of the Fe sitting in position
C mixed with Co); (d)~a fully ordered inverse Heusler structure.
We found, that the structure stability increases from (a) to (d).
The configurations (b) and (c) are isoenergetic, sitting by
285~meV (per formula unit) lower than the ordered regular Heusler structure (a).
Finally, the most stable phase appears to be the  fully ordered inverse
Heusler phase by about 30~meV lower energy compared to (b) and (c). 
Thus, the ordered regular $L2_1$ structure is unstable, and although disorder 
increases the stability, the most stable phase has the ordered inverse
Heusler structure.

This is in good agreement with the experimental results, which suggest that 
Fe$_2$CoGe crystallizes as an inverse Heusler compound with a rather small amount of antisite disorder.
For Fe$_2$CoGe in the X-structure the computed total magnetic moment of
5.15~$\mu_{\rm B}$ is in reasonable agreement with the experimental values of 
5.0-5.4~$\mu_{\rm B}$ (see Table~\ref{tab:magnet} and Ref.~\cite{REN10}).
The calculated local moments are: 2.68, 1.51 and 1.01~$\mu_{\rm B}$ on
Fe(A), Fe(B), and Co, respectively. 
These computed values agree with the Fe moments of 2.6 and 1.6~$\mu_{\rm B}$ derived from the M{\"o}ssbauer 
spectrum of Fe$_{2}$CoGe. The present results are comparable with the calculations of Ref.~\cite{REN10}
which report the moments of 2.74, 1.38 and 0.94~$\mu_{\rm B}$ on Fe(A), Fe(B) and Co 
sites, respectively. By assuming the ferromagnetic order, one derives
from the M\"ossbauer data and the measured magnetization of
5.4~$\mu_{\rm B}$ the local moment of Co as  1.1~$\mu_{\rm  B}$, in  agreement with the present calculations.
%
\begin{figure}[t!]
\centering
\setlength\belowcaptionskip{0.0cm}
\setlength\abovecaptionskip{0.1cm}
\includegraphics[width=0.85\linewidth,clip]{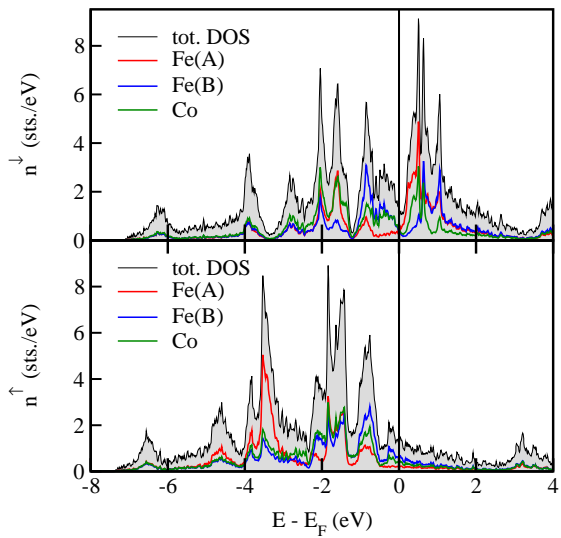}
\caption{
(Color online)
Spin resolved and atom projected DOS of Fe$_2$CoGe, where (a) and (b)
panels correspond to the minority and majority spin channels, respectively.
The red, blue and green curves refer to Fe(A), Fe(B) and Co contributions.
The total DOS is shown by a gray shaded area.
\label{fig:Fe2CoGe-DOS}}
\end{figure}
%
As it follows from the calculated density of states (DOS) for the ordered inverse
Heusler structure (see Fig.~\ref{fig:Fe2CoGe-DOS}) Fe$_2$CoGe is not a
half-metallic ferromagnet (which is also clear from Ref.~\cite{REN10}). 

\subsection{Fe$_2$NiGe}
The EDX data (see Table~\ref{tab_EDX}) for Fe$_2$NiGe indicate that the
sample is slightly off-stoichiometric, with a small portion of Fe 
(about 2.5~\%) is substituted by Ni. The lineshape of the M{\"o}ssbauer spectrum suggests that the sample 
essentially remains ordered, but in the structure there is a certain amount of 
disorder present. So, for Fe$_2$NiGe we have theoretically investigated nine different scenarios
in the inverse Heusler structure.
First we have looked at the stoichiometric compound, where we considered:
(a)~the fully ordered inverse Heusler structure;
(b)~anti-site disorder between Fe(A) and Ni, where 6~\% of Fe was substituted with Ni;
(c)~anti-site disorder between Fe(B) and Ni, where 6~\% of Fe was substituted with Ni;
(d)~6~\% substitution of Fe(A) with Ge;
(e)~6~\% substitution of Fe(B) with Ge.
We have also studied the off-stoichiometric compositions, where the
model systems were created via:
(f)~removal of 3~\% of Fe from site A;
(g)~removal of 3~\% of Fe from site B;
(h)~3~\% substitution of Fe(A) by  Ni; 
(i)~3~\% sustitution of Fe(B) by  Ni. 

Indeed, calculations  indicate that the fully ordered inverse Heusler structure 
is not the most stable configuration for Fe$_2$NiGe.  Instead, the
Fe-Ni anti-site disorder leads to more stable structures. Namely, the
compositions (b) and (c) are lower in energy compared to configuration
(a) by  6 and 8~meV per formula unit, respectively.
In contrast to the Fe-Ni, the Fe-Ge exchange is energetically
unfavorable for each of the Fe sites: compositions 
(d) and (e) are higher in energy by about 65 and 150~meV per formula unit 
compared to the fully ordered case.

Still, these small energy differences between subsequent structures are comparable
with the average thermal energy of the atoms at the annealing temperature,
indicating that the detailed annealing conditions can control the nature of
the disorder in the Fe$_2$YZ inverse Heusler phases.
The energy of configuration (f) is found to be 15~meV lower than (g), indicating that
the creation of Fe point defects in position (A) is more likely than in 
position (B). 

\begin{figure}[t!]
\centering
\setlength\belowcaptionskip{0.0cm}
\setlength\abovecaptionskip{0.1cm}
\includegraphics[width=0.85\linewidth,clip]{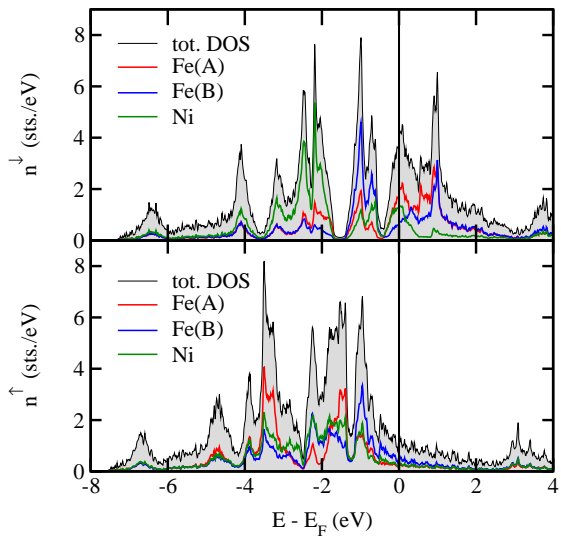}
\caption{
(Color online)
Spin resolved and atom projected DOS of Fe$_2$NiGe, where panels (a) and (b) 
correspond to the minority and majority spin channels, respectively.
The contributions of Fe(A), Fe(B) and Ni are represented with red, blue and
green curves.
The total DOS is shown by a gray shaded area.
\label{fig:Fe2NiGe-DOS}}
\end{figure}

If 3\% of Fe is substituted by an additional 3\% of Ni, the structures
(h) and (i) are isoenergetic, 
i.\,e.\@  in this specific case from the perspective of the total energy the system is 
not biased towards Fe(A) or Fe(B) sites. This means, that with slight Ni
excess, Fe atoms are randomly removed both from positions A and B, and the empty sites 
are being occupied with Ni.  The spin-resolved and atom-projected DOS computed for the fully ordered
stoichiometric Fe$_2$NiGe in the inverse Heusler structure is shown in 
Fig.~\ref{fig:Fe2NiGe-DOS}.
%
%
For the inverse Heusler phase the  calculated total moment is 4.47~$\mu_{\rm B}$
per formula unit, which  agrees  with the experimental moments of 4.29 or 4.46~$\mu_{\rm B}$ measured for the 
samples with SQUID magnetometry. The computed local moments on Fe(A),
Fe(B) and Ni atoms are 2.53, 1.67 and  0.29~$\mu_{\rm B}$ respectively.
The Fe local moments agree with those derived from M\"ossbauer
measurements: 2.5 and 1.7~$\mu_{\rm B}$  for the Fe(A) and Fe(B) sites.   
Similar behavior was found for the disordered structures (b)--(e) corresponding 
to the stoichiometric composition. 

In cases (h) and (i) in which Fe was substituted by Ni, corresponding to the composition measured via EDX, the calculated total
moments were 4.39 and 4.42~$\mu_{\rm B}$, respectively. For structures (f) and (g), where 3~\% of Fe is missing (no additional
Ni present), the  total moments were 4.38 and 4.43~$\mu_{\rm B}$,
respectively. For those cases which resemble the experimental composition, the respective 
local moments were 2.52, 1.68 and 0.31~$\mu_{\rm B}$ for configuration
(i) and for structure (h) 2.54, 1.67 and  0.29~$\mu_{\rm B}$ for Fe(A),
Fe(B) and Ni, respectively. 

Thus, a small amount of disorder, Fe deficiency or
Ni excess do not change considerably the magnetic properties of Fe$_2$NiGe in
the inverse Heusler structure.

\subsection{Fe$_2$NiGa}

In case of Fe$_2$NiGa the experimental results presented in the
previous section have pointed out that the material exhibits rather
strong degree of  disorder. To sample the potential atomic arrangement
in the  disordered phase, we have carried out calculations on eleven model systems
each having a stoichiometric composition.

The first scenario was the ordered inverse Heusler structure, which served as reference.
For the fully ordered structure we have calculated a magnetic moment of 
4.81~$\mu_{\rm B}$, which is by about 0.6~$\mu_{\rm B}$ higher than the experimentally measured
value of 4.20~$\mu_{\rm B}$.

Next we have looked into the effect of antisite disorder 
upon the magnetization and energetics of Fe$_2$NiGa between Fe(A) and Fe(B)
and Ni and Ga for stoichiometric compositions. 
We found, that by exchanging the site occupation of Fe(A) with 10, 20 
and 50\% of Ni, the structures generated hereby are higher in energy by
about 14, 15 and 89~meV per formula unit, respectively, compared to
the ideal inverse Heusler phase. This indicates that a small amount of
Fe(A) (up to about 10--20~\%) can be exchanged by Ni rather easily.

In contrast to this, if one exchanges 50-60~\% of
Fe(B) with Ni, the structure with Fe(B)-Ni antisite disorder becomes 
more stable by about 9~meV per formula unit than the ordered phase.
Thus, in case of Fe$_2$NiGa the ground state does not correspond to the
fully ordered inverse Heusler phase.
For the lowest energy configuration  the calculated total moment is
4.77~$\mu_{\rm B}$, where the local moments on Fe(A), Fe(B) and Ni are
2.65, 1.86 and  0.33~$\mu_{\rm B}$, respectively. Thus, the magnetic
moment of this structure  is still higher 
than the SQUID data by about 4.2~$\mu_{\rm B}$ per unit cell. 

Also the average moment on the Fe sites of 2.25~$\mu_{\rm B}$ is
somewhat higher than the average moment of 2.05~$\mu_{\rm B}$ estimated from the broad sextets in the M{\"o}ssbauer spectrum.
Upon exchanging either Fe(A) or Fe(B) with Ga the disordered structures were all
unstable relative to the reference case independent of the Fe position.
The total magnetic moments of the systems with Fe-Ni and Fe-Ga 
exchange are computed in the range from 5.03 to 4.59~$\mu_{\rm B}$.  

Due to the fact that the M{\"o}ssbauer spectra may suggest random
alloying,  we have also considered the case of an A2-type structure, where
the constituent atoms are randomly distributed on all four lattice sites.
The calculations have shown, that such an A2-type random structure where all 
atoms are coupled ferromagnetically, is rather unstable, by about 440~meV per formula unit
compared to the reference, which is the  ordered inverse Heusler phase.

\subsection{Fe$_2$CuGa}

\begin{figure}[t!]
\centering
\setlength\belowcaptionskip{0.0cm}
\setlength\abovecaptionskip{0.1cm}
\includegraphics[width=0.85\linewidth,clip]{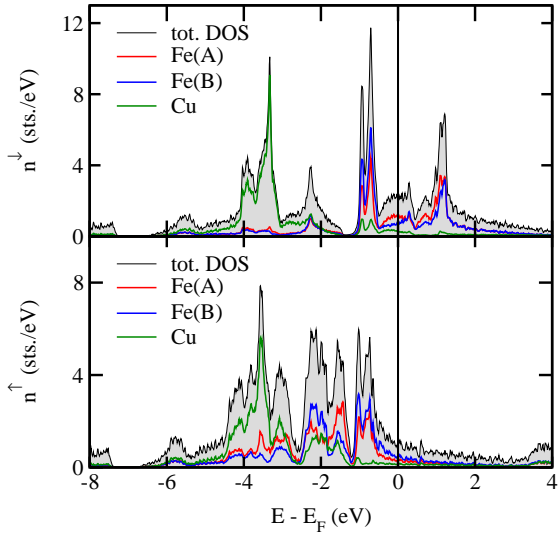}
\caption{
(Color online)
Spin resolved and atom projected DOS of Fe$_2$CuGa.
Panels (a) and (b) correspond to the minority and majority spin channels, respectively.
The contributions of Fe(A), Fe(B) and Cu are shown with 
red, blue and green curves, respectively, and the total DOS is shown by a gray shaded area.
\label{fig:Fe2CuGa-DOS}}
\end{figure}

For the Fe$_2$CuGa sample the experiment reveals both well-ordered and disordered regions.
Since the ordered regions should correspond to the ideal ferromagnetic inverse 
Heusler structure, we have first investigated this configuration.
The corresponding spin-resolved and atom-projected DOS is shown in Figure~\ref{fig:Fe2CuGa-DOS}.
%
%
The calculated total magnetic moment of 4.04~$\mu_{\rm B}$ is somewhat larger than the
 experimental one (3.6~$\mu_{\rm B}$) derived from the SQUID measurements.
The origin of this small discrepancy cannot be resolved unambiguously, however,
the experimental moment may be too small due to the presence of the minor impurity phase 
as suggested by the XRD pattern.  On the other hand, the calculated
local moments of 2.34~$\mu_{\rm B}$ for Fe(A) and
1.78~$\mu_{\rm B}$ for Fe(B) sites  agree with the corresponding values of 
2.5 and 1.7~$\mu_{\rm B}$ derived from the sharp features in the M{\"o}ssbauer spectrum. 
The moments on the Cu and Ga sites are negligibly small.

To gain new insights via CPA on the potential nature of the disordered phase, 
we have carried out calculations for six configurations taking into account
various kinds of Fe-Cu and Fe-Ga antisite disorder.
The calculations indicate that 50~\% of Cu-Fe(A) exchange 
leads to a more stable structure compared to the fully ordered inverse Heusler 
structure with a stabilization energy of 71~meV per primitive unit cell. 
Beside this, we found that an even more stable configuration can be created
via chemical disorder, where alloying is present between Fe and Cu, with 
18, 44 and 44~\% of Fe-Cu antisite exchange  on Wyckoff positions A, B and D,
respectively. This structure is by 110~meV lower in energy compared to the ideal ordered inverse
Heusler structure, which indicates that the system might undergo
a spontaneous mixing between Fe and Cu, giving rise to the appearance
of a broad sextet seen in  the M{\"o}ssbauer spectra.
For the systems with Fe-Cu disorder the total magnetic moments assuming 
ferromagnetic ordering have been calculated to be in the range of
4.29-4.31~$\mu_{\rm B}$, where the local moments were 2.48, 1.95 and
1.94~$\mu_{\rm B}$ for Fe(A), Fe(B) and Fe(D), respectively. The average
value of 2.1~$\mu_{\rm B}$ over these sites is in good agreement with
the average moment of 2.1~$\mu_{\rm B}$ estimated from the hyperfine field value of the broad sextet.

Next we considered the effect of Fe-Ga disorder, which has shown 
that even a small amount of Fe-Ga antisites is  energetically unfavorable.
For 10~\% of antisite disorder on Fe(A) or Fe(B) sites the structures were less
stable, by about  30 and 85~meV  per formula unit, respectively, compared to 
the ordered inverse Heusler structure. 
This indicates that Fe-Ga antisite disorder in Fe$_2$CuGa is less
likely to happen than Fe-Cu alloying. The calculated total moments in
case of  Fe-Ga antisite disorder are in the range of
4.15-4.25~$\mu_{\rm B}$ which are only slightly enhanced compared to the ideal X-structure.

\subsection{Fe$_2$CuAl}
\begin{figure}[t!]
\centering
\setlength\belowcaptionskip{0.0cm}
\setlength\abovecaptionskip{0.1cm}
\includegraphics[width=0.85\linewidth,clip]{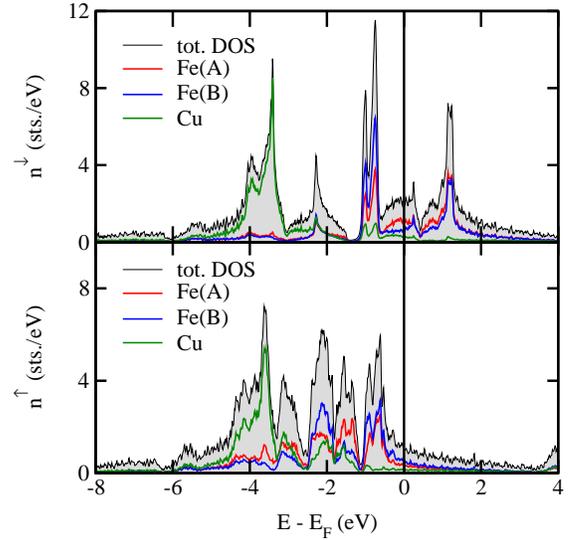}
\caption{
(Color online)
Spin resolved and atom projected DOS of Fe$_2$CuAl, where (a) and (b)
panels correspond to the minority and majority spin channels, respectively.
The red, blue and green curves refer to Fe(A), Fe(B) and Cu contributions.
The total DOS is shown by a gray shaded area.
\label{fig:Fe2CuAl-DOS}}
\end{figure}

For the Fe$_2$CuAl sample it was found experimentally, that the sextet assigned
to Fe(B) in the M{\"o}ssbauer spectra has a lower intensity than it should have in the ordered
inverse Heusler phase, indicating that there is a certain amount
of disorder involved. Also, there is a second broad sextet with enhanced relative intensity present, pointing towards different local 
Fe coordination environments. These spectral features suggest an inverse Heusler 
structure, where Fe(B)-Al antisite disorder leads to an enhanced fraction of Fe(A)-like sites. 
We have first investigated the ideal X-phase of Fe$_2$CuAl theoretically, for
which the computed total DOS and atom projected DOS is shown in
Fig.\ref{fig:Fe2CuAl-DOS}.
%
%
The calculated total moment of 3.80~$\mu_{\rm B}$ considering ferromagnetic ordering 
is somewhat higher than the measured value of 3.3~$\mu_{\rm B}$.  The
experimental value may, however, be too small as there is an impurity 
phase evident in the XRD pattern. Similar to  the case of Fe$_2$CuGa the
local moments deduced from the M{\"o}ssbauer spectra on Fe(A) and Fe(B) sites of 2.2 and 1.7~$\mu_{\rm B}$, 
are in a real good agreement with the calculated local moments of 2.23 and 1.67~$\mu_{\rm B}$. 
This confirms that Fe$_2$CuAl basically crystallizes in a cubic inverse Heusler structure.
Further, we studied the effects of  Fe-Cu or Fe-Al antisites.
Up to 30\% of Fe-Cu antisite exchange the structure is more stable by
about 12~meV per formula unit compared to the ideal inverse structure, and going above 30\% the
structure become less stable by about around 15~meV.

By exchanging Fe with Al the calculations show that the structures
generated are practically isoenergetic with the ordered inverse Heusler
structure.
This result agrees with the experimental observations from
Ref.~\onlinecite{VBC09}. The total moments for configurations with Fe-Cu
and Fe-Al disorder are in the range of 3.91-3.96 and 3.80-3.81~$\mu_{\rm B}$, 
respectively, which is similar as in case of the fully ordered Heusler phase.
Thus, the calculations indicate, that although there is a slight bias
towards the Fe-Cu antisite disorder, in Fe$_2$CuAl the spontaneous formation of
both Fe-Cu and Fe-Al type disorder is highly likely. As it follows, in
many cases  the antisite disorder does not change radically the magnetic
properties of the Heusler compounds.\\
\indent

\subsection{Fe$_2$ZnAl}
Finally we considered the case of Fe$_2$ZnAl (predicted
theoretically~\cite{GD09}).  We have investigated the hypothetical
inverse cubic Heusler structure, and found that the calculated total
moment of 3.27~$\mu_{\rm B}$ is nearly twice larger
than the experimental moment of a sample with the nominal composition
Fe$_2$ZnAl. Such significant discrepancy suggests that the sample is not a Heusler
phase. In fact, the XRD and M{\"o}ssbauer data revealed that in the mixed-phase samples 
non-magnetic FeAl, in addition to Fe$_3$Al were the major iron-based components.

\section{Conclusions}
We have synthesized and characterized the new iron-based Heusler phases 
Fe$_{2}$NiGe, Fe$_{2}$CuGa, and Fe$_{2}$CuAl, the existence of which has been 
suggested by previous theoretical work~\cite{GD09}. 
In addition we have investigated in more detail the known compounds 
Fe$_{2}$CoGe and Fe$_{2}$NiGa. 
The newly synthesized Fe$_{2}$YZ phases were predicted to adopt a regular tetragonal
Heusler structure, whereas the present experimental results revealed that the 
materials basically crystallize in the cubic inverse Heusler (X-) structure with
differing degrees of atomic disorder. 
For these Fe-based materials $^{57}$Fe M{\"o}ssbauer spectroscopy is a very useful
local-probe technique to unravel the atomic disorder and to estimate the magnetic
moments at the iron sites. 
All the compounds are soft ferromagnets with Curie temperatures up to about 900 K 
which makes them suitable for potential magnetic applications. 
The electronic structures of the materials have been studied by {\it ab-initio} DFT 
calculations including the effects of disorder and nonstoichiometry within the CPA 
approximation. 
The good agreement between calculated and experimental magnetic moments at the Fe 
sites is further support that the compounds basically adopt the cubic inverse 
Heusler structure. 
Atomic disorder which is apparent in the experimental data leads to further 
stabilization of the cubic structure without changing the magnetic properties too much. 
The case of Fe$_{2}$NiGa, which was predicted to crystallize in the cubic X-structure 
is not completely clear yet. 
Both, XRD and M{\"o}ssbauer data evidence an increased atomic disorder and the 
calculated magnetic moments are somewhat higher than the experimental ones. 
Attempts to synthesize Fe$_{2}$ZnAl, a further phase which was also suggested 
to adopt the tetragonal Heusler structure~\cite{GD09}, were unsuccessful.
While there is a good consistency between the present experimental and theoretical 
results for the cubic inverse Heusler Fe$_{2}$YZ materials it remains to be clarified
why Fe$_{2}$NiGe, Fe$_{2}$CuGa, and Fe$_{2}$CuAl do not adopt the tetragonal structures
predicted previously. 

\bigskip
\begin{acknowledgments}

We thank Dr. G\"ostar Klingelh\"ofer for the possibility to use the MIMOS spectrometer, 
Dr. Mathias Blumers for his support, and Dr. Gerhard H. Fecher for helpful discussions. 
Financial support by the $Deutsche Forschungsgemeinschaft$ (project P2.3-A of research unit
FOR 1464 $ASPIMATT$) is greatfully acknowledged.   

\end{acknowledgments}

%

\end{document}